\documentclass{PoS}
\usepackage{cite}

\title{Search for the  eta-mesic helium by means of WASA detector at COSY}

\ShortTitle{Search for the eta-mesic helium}

\author{\speaker{P. Moskal} \\ 
        M. Smoluchowski Institute of Physics, Jagiellonian University, 30-059 Cracow, Poland,\\
        IKP, Forschungszentrum J\"ulich, D-52425 J\"ulich, Germany\\
        E-mail: \email{p.moskal@uj.edu.pl}}

\author{W. Krzemie{\'n} \\ 
        M. Smoluchowski Institute of Physics, Jagiellonian University, 30-059 Cracow, Poland,\\
        E-mail: \email{wojciech.krzemien@if.uj.edu.pl}}

\author{M. Skurzok \\ 
        M. Smoluchowski Institute of Physics, Jagiellonian University, 30-059 Cracow, Poland,\\
        E-mail: \email{magdalena.skurzok@uj.edu.pl}}

\author{for the WASA-at-COSY Collaboration}  

\abstract{
  The $\eta$-mesic nuclei in which the $\eta$ meson is bound 
  with nucleus via strong interaction was postulated 
  about 25 years ago, however till now no experiment
  confirmed empirically its existence. The discovery 
  of this new kind of an exotic nuclear matter would 
  be very important for better understanding of the 
  $\eta$ meson structure and its interaction with nucleons. 
  The search for $\eta$-mesic helium is carried out with high
  statistic and high acceptance with the WASA-at-COSY
  detection setup and the Cooler Synchrotron COSY. 
  The search is conducted via the measurement of the
  excitation function for the chosen decay channels of the
  $^4\!He-\eta$ system. Until now two reactions 
  $dd \to (^4\!He-\eta)_{bound} \to ^3\!He p \pi^-$ 
  and 
  $dd \to (^4\!He-\eta)_{bound} \to ^3\!He n \pi^0$
 have been measured with the beam momentum varied continuously
  around the $\eta$ meson production threshold. This contribution
  includes description of recently published results of the WASA-at-COSY experiment
  as well as brief presentation of preliminary results from 
  the analysis of new data sample with 20 times larger
  statistics. 
          }

\FullConference{XV International Conference on Hadron Spectroscopy-Hadron 2013\\
		4-8 November 2013\\
		Nara, Japan }

\begin{document}

\section{Introduction}
The motivation for the search of the mesic nuclei~\cite{Haider} was described in many articles 
and for the newest results the interested reader is referred to contributions from the recent 
symposium dedicated to the mesic nuclei
e.g.~\cite{strona,Mesic-Bass,Colin-ACTA,GAL-ACTA,KruscheActa,Wycech-Acta}.
Therefore, here we only briefly recall that
the  observation of a bound state of meson and nucleus would not only be interesting on its own account,
but the determination of properties of $\eta$-mesic 
or $\eta^{\prime}$-mesic nucleus would be valuable for 
establishing of the $\eta$ and $\eta^{\prime}$ interactions with nucleons,
for studies of the N$^*(1535)$ properties in nuclear matter~\cite{jido,EtaMesic-Hirenzaki},
the properties of the $\eta$ and $\eta^{\prime}$ mesons
in the nuclear medium~\cite{osetNP710,ETA-Friedman-Gal,WYCECHGREE-GW-zGAla},
for the determination of the flavour singlet components
of $\eta$ and $\eta^{\prime}$ mesons~\cite{GLUE-PL-Bass06},
and in general for studies of the chiral and axial U(1) symmetry breaking
in low energy QCD~\cite{EtaMesic-Hirenzaki,GLUE-PL-Bass06,Mesic-Bass}.

At present, there are ongoing research programs on the $\eta$ and $\eta'$-mesic nuclei
     in many laboratories as e.g.:
  COSY~\cite{COSY11-MoskalSmyrski,WASA-at-COSY-SkuMosKrze,Adlarson2013,GEM,3Heeta-PL-Smyrski},
  ELSA~\cite{Nanova},
  GSI~\cite{eta-prime-mesic-GSI-tanaka},
  JINR~\cite{JINR},
  JPARC~\cite{eta-mesic-JPARC-Fujioka},
  LPI~\cite{LPI}, and
  MAMI~\cite{ELSA-MAMI-plan-Krusche,gamma3He-Pheron}. The experimental endeavour to discover the mesic nucleus is supported by the intensive
theoretical
investigations e.g.~\cite{
Mesic-Bass,
Wycech-Acta,
EtaMesic-Hirenzaki,
ETA-Friedman-Gal,
WYCECHGREE-GW-zGAla,
GLUE-PL-Bass06,
wilkin2,
eta-prime-mesic-Nagahiro,
eta-prime-mesic-Hirenzaki,
eta-prime-mesic-Nagahiro-Oset,
ETA-Gal-Cieply,
Mesic-Kelkar,
Bass10Acta}.
Search for the direct signal from the $\eta$-mesic nucleus are being continued since many years,  
however so far none of the experiments delivered an unambiguous
signal, which could confirm existence of such object.
In the case of the production of the $\eta$-mesic helium the established upper limits amount to
about 270~nb
for the
 $dp \to ({^{3}\mbox{He}} \eta)_{bound} \to ppp\pi^-$  reaction~\cite{COSY11-MoskalSmyrski},
about 70~nb
for the
 $dp \to ({^{3}\mbox{He}} \eta)_{bound} \to {^{3}\mbox{He}} \pi^0$  reaction~\cite{COSY11-MoskalSmyrski},
and
about 25~nb
for the
 $dd \to ({^{4}\mbox{He}} \eta)_{bound} \to {^{3}\mbox{He}} p\pi^-$  reaction~\cite{Adlarson2013}.
The achieved experimental sensitivity is close to the newly predicted values of total cross sections
amounting to about 80~nb for the $pd \to ({^{3}\mbox{He}}-\eta)_{bound} \to X p \pi^-$~\cite{Colin-ACTA}
and 4.5~nb~\cite{Wycech-Acta} to
30~nb~\cite{Colin-ACTA} for the $dd \to (^4\!He\eta)_{bound} \to X p \pi^-$ reaction.

There are, however, very encouraging observations indicating indirectly the existence of the $\eta$-mesic helium, 
which can be summarized as follows:
  {\bf (1)} 
  The extremely steep rise of the total cross section for the $pd\to{^{3}\mbox{He}}\eta$ reaction
    in the very close-to-threshold region followed
    by a plateau~\cite{3Heeta,3Heeta-PL-Smyrski}
    suggesting a pole
    in the $\eta {^{3}\mbox{He}}\to \eta {^{3}\mbox{He}}$ scattering amplitude at the complex excess energy
    plane $Q$ with $Im(Q)<0$~\cite{wilkin2}. 
    A large enhancement of the total cross section at threshold is observed also for the reactions 
    $dd \to ^4\!\!He \eta$~\cite{4Heeta} and
    $pd \to  pd \eta$~\cite{pdeta}; 
  {\bf (2)} 
    A steep increase of the total cross section for ${^{3}\mbox{He}}-\eta$ photo-production at threshold
    via the $\gamma ^3He \to \eta ^3He$ reaction~\cite{KruscheActa,gamma3He-Pheron} 
    shows that the rise of the cross section above threshold is independent of the initial channel
    and can therefore be assigned to the $^3He-\eta$ interaction;
  {\bf (3)} 
    The recent determination of the energy dependence of the tensor analysing power $t_{20}$ by the ANKE collaboration
    confirmed that the s-wave production amplitude in the $pd \to{^{3}\mbox{He}}\,\eta$ reaction
    is fairly energy independent~\cite{3Heeta-t20-Khoukaz} 
    again indicating that the steep threshold
    enhancement is due to the $^3He-\eta$ interaction;
  {\bf (4)} 
    The asymmetry in the angular distribution of the $\eta$ meson emission~\cite{3Heeta-PL-Smyrski}
    indicates strong changes of the phase of the s-wave production amplitude with energy,
    as expected from the occurrence of the
    bound or virtual $\eta {^{3}\mbox{He}}$ state~\cite{wilkin2}; 
  {\bf (5)} 
    The evolution with energy of the angular dependence of~$\gamma ^3He \to \eta ^3He$~\cite{gamma3He-Pheron}
    is "similar to that of the $pd\to {^{3}\mbox{He}}\eta$ reaction which indicates changing of s-wave amplitude
    associated with the $\eta{^{3}\mbox{He}}$ pole"~\cite{Colin-ACTA}.

    In addition, it is worth to stress that an often stated argument that the extracted $\eta N$ scattering 
    length is too low for the $\eta$-helium binding is weakened in view of 
    new theoretical results. Estimates of sub-threshold amplitudes are model dependent and recently Gal et al., has concluded that:
    "Calculations of $\eta$-nuclear bound states show, in particular, that the $\eta$-N scattering length is not a useful indicator
    of whether or not $\eta$ meson bind in nuclei"\cite{GAL-ACTA}.
    Moreover, differences in the value of $\eta N$ scattering lengths obtained in different analyses are at least to some extent
    explained by the recent observation that the flavour singlet component induces greater binding than the flavour-octet one.
    Therefore, the $\eta-\eta^{\prime}$ mixing, which is neglected in many of the former analyses,
    increases the $\eta$-nucleon scattering length relative
    to the pure octet $\eta$ by a factor of about 2~\cite{Mesic-Bass}.

The binding energies of $\eta$ and $\eta^{\prime}$ 
in nuclear medium are sensitive to the non-perturbative glue~\cite{GLUE-PL-Bass06,Mesic-Bass}
and due to
the UA(1) anomaly effect, a relatively large mass reduction of $\eta^{\prime}$ meson is expected in medium~\cite{Hirenzaki-ACTA} 
which may indicate the existence of the $\eta^{\prime}$-mesic nucleus~\cite{Hirenzaki-ACTA} as predicted  
in~\cite{eta-prime-mesic-Hirenzaki}. Therefore, in spite of the indications that 
$\eta^{\prime}$-nucleon interactions is smaller than 
$\eta$-nucleon~\cite{Moskal-Swave},
recently there are vigourous
theoretical investigations~\cite{Mesic-Bass,EtaMesic-Hirenzaki,eta-prime-mesic-Nagahiro,eta-prime-mesic-Nagahiro-Oset,eta-prime-mesic-Nagahiro-Jido}
and preparations of experiments for the $\eta^{\prime}$-mesic nucleus~\cite{ELAS-BGO-OD-Metag,eta-prime-mesic-GSI-tanaka}.
These experiments are motivated also by
the recent
photoproduction measurements of CBELSA/TAPS~\cite{Nanova}
showing that the
real part of the $\eta^{\prime}$-nucleus optical potential is larger than  its imaginary part.

\section{Status of the search for $\eta$-mesic helium with WASA-at-COSY}

The WASA detector~\cite{Adam} at COSY gives unique possibilities to conduct
 studies of the hadronic production of ${\mbox{He}}-\eta$ system
     with the continuous change of the beam momentum and the exclusive measurement of all ejectiles.
Two experiments dedicated to the search of $\eta$-mesic helium were conducted up to now using the WASA-at-COSY detector.
Both were focused on the bound state decay into the ${^{3}\mbox{He}}$ and a nucleon-pion pair~\cite{WASA-at-COSY-SkuMosKrze,WojciechPoS}

The first experiment was performed in June 2008 by measuring the excitation function 
of the $dd \rightarrow$ $^{3}\hspace{-0.03cm}\mbox{He} p \pi^{-}$  reaction near the $\eta$ meson production threshold~\cite{Adlarson2013}.
During the experimental run the momentum of the deuteron beam was varied continuously within each acceleration cycle
from  2.185~GeV/c to 2.400~GeV/c, crossing the kinematic threshold for $\eta$ production in the $dd \rightarrow ^4$He$\,\eta$ reaction at 2.336~GeV/c.
This range of beam momenta corresponds to a variation of the $^4\mbox{He}-\eta$  excess energy  from -51.4~MeV to 22~MeV.
The data show no statistically significant signal of the $^4\mbox{He}$-$\eta$ bound state in the excitation function~\cite{Adlarson2013}, and the 
determined 
upper limit for the cross-section for the bound state formation and decay in the process
$dd \rightarrow ({{^4\mbox{He}}-\eta})_{bound} \rightarrow ^3$He$ p \pi^{-}$
varies from 20~nb to 27~nb at 90\%CL~\cite{Adlarson2013}. 
The established upper limits depend mainly on the width of the bound state and only slightly on the binding energy.

In the second experiment, conducted in November 2010 we increased the statistics with respect to the previous results~\cite{Adlarson2013}
by a factor of about 20.
During the second experimental campaign two reactions were measured simultaneously:
$dd\rightarrow(^{4}\mbox{He}$-$\eta)_{bound}\rightarrow$ $^{3}\mbox{He} p \pi{}^{-}$ 
and  
$dd\rightarrow(^{4}\mbox{He}$-$\eta)_{bound}\rightarrow$ $^{3}\mbox{He} n \pi{}^{0} \rightarrow$ $^{3}\hspace{-0.03cm}\mbox{He} n \gamma \gamma$~\cite{WASA-at-COSY-SkuMosKrze}.
The deuteron beam momentum was varied continuously within each acceleration cycle 
in the momentum range corresponding to the 
variation of the $^4\mbox{He}$-$\eta$ excess energy  from -70~MeV to 30~MeV.
The analysis of the data is still in progress. 
Preliminary spectra of the excitation function have been presented in~\cite{WojciechPoS}.
The shape of the excitation functions determined for the range of small $^3\!He$ momenum, when a signal from the bound state is expected, differes from the shape
of the excitation function for the range with larger $^3\!He$ momenum. This is a promising result, but 
its final interpretation requires detailed simulations in order to understand the 
background contributions to the observed excitation functions.  

Anyhow, already now we can conclude that a collected data are of
a very good quality and that we have achieved a sensitivity
of the cross section of the order of few nb for the bound state
production in $dd\rightarrow$ $(^{4}\hspace{-0.03cm}\mbox{He}$-$\eta)_{bound} \rightarrow$ $^{3}\hspace{-0.03cm}\mbox{He} n \pi^{0}$
and $dd\rightarrow$ $(^{4}\hspace{-0.03cm}\mbox{He}$-$\eta)_{bound} \rightarrow$ $^{3}\hspace{-0.03cm}\mbox{He} p \pi^{-}$ reactions.
This is in the range of cross section value of 4.5~nb predicted in~\cite{Wycech-Acta}.

\section{Perspectives}

For possible future investigations of the $\eta$-mesic helium an interesting
possibility of non-$N\pi$ decays of the mesic-helium
was pointed out by Wycech~\cite{Wycech-Acta} and Wilkin~\cite{Colin-ACTA}.
In this case the reaction may proceed e.g. as follows:
    $pd\to ({^{3}\mbox{He}}-\eta)_{bound}\to ppn$ or
    $pd\to ({^{3}\mbox{He}}-\eta)_{bound}\to pd$.
Such processes could be due to the absorption of $\eta$-meson via e.g. $\eta d \to pn$ reaction. However, Wilkin~\cite{Colin-ACTA}
 estimated in the first approximation
that the two-nucleon absorption constitutes at most 5\% of the total decay rate~\cite{Colin-ACTA},
and in addition these channels are buried in a large background.

Another possible mechanism of the decay of the bound state 
is the decay of the $\eta$ meson when it is still orbiting around the nucleus.
This seems to be promising experimentally due to the very low background.
Using arguments given in~\cite{Colin-ACTA},
as a very rough approximation we may estimate the cross sections for the processes:
$pd \to ({^{3}\mbox{He}}$-$\eta)_{bound} \to {^{3}\mbox{He}}\,2\gamma$, and
$pd \to ({^{3}\mbox{He}}$-$\eta)_{bound} \to {^{3}\mbox{He}}\,6\gamma$
to be about 0.4~nb. This value can be estimated taking into account that the total width 
of the $\eta$ meson is about 1.3 keV, the width of the (${^{3}\mbox{He}}-\eta$)
is less than about 500 keV, and the $2\gamma$ and $6\gamma$ branching ratios amounts
to about 39\% and 33\%, respectively~\cite{Colin-ACTA}.

\section{Acknowledgements}
We acknowledge support
by the Foundation for Polish Science (MPD programme),
by the Polish National Science Center through grant: 
No. 
2011/01/B/ST2/00431, 
and by the FFE grants of the Research Center J\"{u}lich.

\end{document}